\pdfoutput=1

\documentclass[11pt]{article}

\usepackage{CJKutf8}
\usepackage{acl}
\usepackage{amsmath}
\usepackage{graphicx}
\usepackage{times}
\usepackage{latexsym}
\usepackage{enumerate}
\usepackage{algorithm}
\usepackage{enumitem}
\usepackage[T1]{fontenc}
\usepackage{multirow}
\usepackage{booktabs}
\usepackage[utf8]{inputenc}
\usepackage{CJKutf8} 
\usepackage{makecell}
\usepackage{microtype}
\usepackage{graphicx}
\usepackage{epstopdf}

\title{Event-Centric Query Expansion in Web Search}

\author{Yanan Zhang\textsuperscript{$1$}\thanks{\;\;Equal Contributions}  \enskip Weijie Cui\textsuperscript{$2*$} \enskip Yangfan Zhang\textsuperscript{$1$} \enskip Xiaoling Bai\textsuperscript{$1$}\thanks{\;\;Corresponding author} \\
\textbf{Zhe Zhang}\textsuperscript{$2$} \enskip \textbf{Jin Ma}\textsuperscript{$2$} \enskip \textbf{Xiang Chen}\textsuperscript{$1$} \enskip \textbf{Tianhua Zhou}\textsuperscript{$1$} \\
 \textsuperscript{$1$}Tencent Inc. \quad \textsuperscript{$2$}University of Science and Technology of China\\
   \{yananzhang, devinbai\}@tencent.com,  \ \ can@mail.ustc.edu.cn \\
}

\begin{document}
\begin{CJK*}{UTF8}{gbsn} 
\maketitle
\begin{abstract}

In search engines, query expansion~(\textbf{QE}) is a crucial technique to improve search experience. Previous studies often rely on long-term search log mining, which leads to slow updates and is sub-optimal for time-sensitive news searches. In this work, we present \textbf{E}vent-Centric \textbf{Q}uery \textbf{E}xpansion~(EQE), a novel QE system that addresses these issues by mining the best expansion from a significant amount of potential events rapidly and accurately. This system consists of four stages, i.e., \textit{event collection}, \textit{event reformulation}, \textit{semantic retrieval} and \textit{online ranking}. Specifically, we first collect and filter news headlines from websites. Then we propose a generation model that incorporates contrastive learning and prompt-tuning techniques to reformulate these headlines to concise candidates. Additionally, we fine-tune a dual-tower semantic model to function as an encoder for event retrieval and explore a two-stage contrastive training approach to enhance the accuracy of event retrieval. Finally, we rank the retrieved events and select the optimal one as QE, which is then used to improve the retrieval of event-related documents. Through offline analysis and online A/B testing, we observe that the EQE system significantly improves many metrics compared to the baseline. The system has been deployed in Tencent QQ Browser Search and served hundreds of millions of users. The dataset and baseline codes are available at \url{https://open-event-hub.github.io/eqe}.

\end{abstract}

\section{Introduction}
\begin{figure}[t!]
    \centering
    \includegraphics[width=\linewidth]{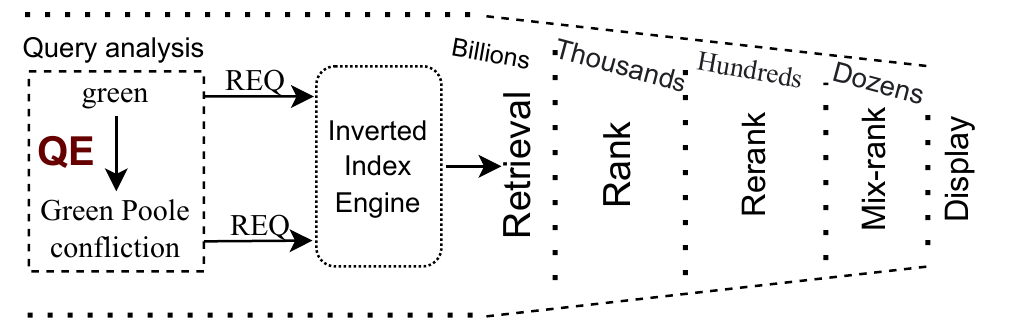}
    \captionsetup{skip=3pt} 
    \caption{Overview of the query expansion process and search system in Tencent QQ Browser Search.}
    \label{fig:overview}
    \vspace{-1.5em}
\end{figure}

People are always eager to obtain details and updates on current hot events through search engines. To efficiently return dozens of relevant documents from billions of candidates, most search engines use a ``retrieval-rank-rerank-mixed rank''architecture, as illustrated in Figure~\ref{fig:overview}.

Queries, particularly those relying on keywords, present a tough challenge for query intent understanding due to their brevity, absence of world knowledge, and lack of grammatical structure~\citep{broder2007robust}. 
When a significant event takes place, the search intent of users subsequently can rapidly and drastically shift. For example, during the Green-Poole conflict, a user searching for ``green'' may be seeking information about the color, while many others desire news about NBA player Draymond Green.
While methods based on search log mining~\citep{jansen2007determining, zamora2014query, caruccio2015understanding} are still commonly used for query intent understanding, they are limited by their reliance on the accumulation of posterior data and struggle with timely and accurately processing the intent for recent events, making it difficult to retrieve and rank event-related documents.
Recent approaches~\citep{zhang2020query, nogueira2019doc2query, Sun2022} suggest using additional context from query-associate documents or entities to improve the performance of query understanding. However, they still face challenges in real-time search scenarios.

\begin{figure*}[ht]
\vspace{-1.0em}
    \centering
    \includegraphics[width=0.95\textwidth]{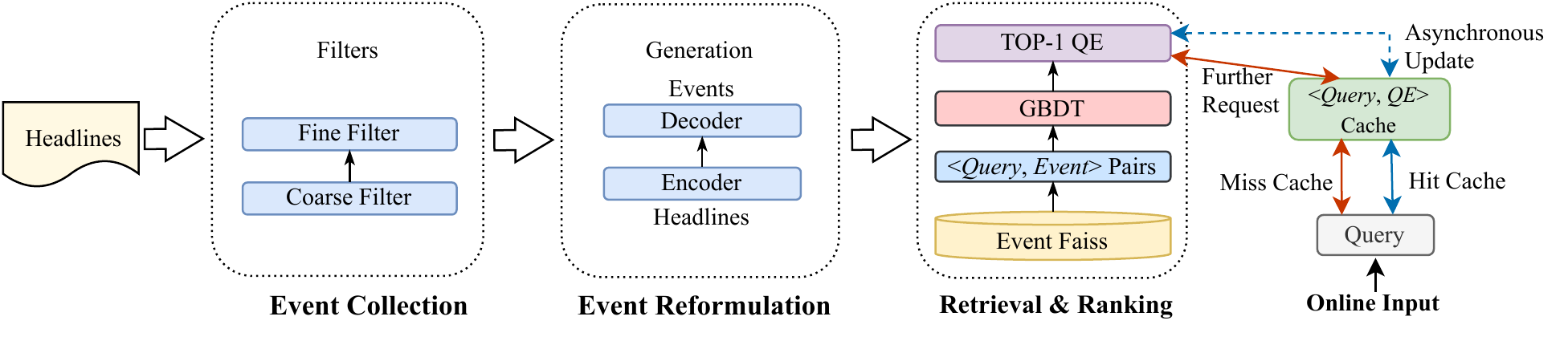}
    \caption{Architecture of the proposed EQE system.}
    \label{fig:system}
\vspace{-0.5em}
\end{figure*}

To tackle this challenge, we present EQE, a real-time query expansion system specifically designed to efficiently capture query intent for ongoing events.
As depicted in Figure~\ref{fig:overview}, EQE extends the original query with the most fulfilling event, selected from a large pool of candidate events. By performing the same retrieval step with both the original and expanded queries, more results related to the event will be returned. This bypass architecture effectively ensures system flexibility. When there are sufficient machine resources, the number of bypasses can be increased, and multiple query expansions can be used to improve the performance of document retrieval.

Our EQE system employs a four-stage structure, as illustrated in Figure~\ref{fig:system}, consisting of event collection, event reformulation, semantic retrieval, and online ranking. 
Events are collected from news headlines as they are typically more concise and event-centric, compared to body texts which are lengthier and contain extraneous information. To guarantee the accuracy of the collected events, we employ a combination of rule-based coarse filtering and language model-based fine filtering~($\S$~\ref{sec:event_collect}).
The collected headlines, as described in Appendix~\ref{sec:CharacteristicsCET}, may contain noise, irregular grammar, and lack of world knowledge, making them unsuitable for query expansions. To solve these problems in such scenarios, we reformulate them using a generation model, which is more effective than extractive models~($\S$~\ref{sec:EventReformulation}). Our method employs keyword-based prompt learning to make generated content more controllable and applies contrastive learning on the encoder to counteract embedding degradation~\citep{gao2019representation}. After this step, we obtain a high-quality candidate set of event-centric QE.
For a given query, to further narrow down the QE candidate set, we utilize a supervised SimCSE model~\citep{gao2021simcse} to retrieve relevant QEs. SimCSE effectively improves the accuracy of retrieval by addressing the issue of representation space degradation. Moreover, inspired by ~\citep{DanielGillick2019LearningDR}, we employ a two-stage training approach that incorporates informative hard negative samples for each query, resulting in a further improvement in representation quality~($\S$~\ref{sec:SemanticRetrieval}).
Finally, we design an online ranking module to select the best QE. Features of query-side, event-side, and interactive are considered comprehensively~($\S$~\ref{sec:online_ranking}).

As far as we know, EQE is the first query expansion solution developed explicitly for real-time event intent.
The efficiency of EQE is verified through offline analysis and online A/B testing. The main contributions of this work are summarized as follows:

\begin{itemize}
    \item We propose a real-time and efficient query expansion system for timely search scenarios. The system comprises four stages: event collection, event reformulation, semantic retrieval, and online ranking.
    \item In the event reformulation stage, we introduce an effective generation model that leverages prompt learning and contrastive learning techniques to produce a high-quality candidate set of QE.
    \item In the semantic retrieval stage, we employ a two-stage contrastive learning approach to improve the accuracy of semantic retrieval.
    \item  Offline analysis and online A/B testing on Tencent QQ Browser Search demonstrate the effectiveness of our proposed EQE framework.
\end{itemize}

\section{Method}
In this section, we describe our proposed framework of EQE shown in Figure~\ref{fig:system}. 
We first introduce the scheme for event collection in industrial scenarios. 
We then elaborate on event reformulation and semantic retrieval, describing how we use contrastive learning and prompt learning to improve model performance. 
Finally, we discuss online ranking, revealing how to select the optimal expansion.

\subsection{Event Collection}\label{sec:event_collect}

The essential phase in the ``Event Collection'' process is to identify events from the vast amount of newly uploaded content on the Internet. We filter events from the headlines using a two-step method that includes a rule-based coarse filter and a semantically-driven fine filter.

\noindent \textbf{Coarse Filter.} 
After using basic feature filters such as publication time, page type, site type, etc., we gather approximately 50 million news article headlines over the duration of recent six months.
As described in Appendix~\ref{sec:event_fi}, the headlines generated by these heuristic rules include irregular syntax, missing event components, numerous events, etc., posing a barrier to subsequent event reformulation.
So further we use the LTP toolkit~\citep{che-etal-2021-n} to extract event triggers from headlines and drop headlines missing event elements or with multiple events~(the number of triggers more than 2). 

\noindent \textbf{Fine Filter.} 
The rule-based coarse filter is based on pre-defined patterns and has limited recognition abilities. To address these issues and further improve the accuracy of event collection, we train an event detection model based on RoBERTa~\citep{YinhanLiu2019RoBERTaAR}.
We employ six experts to annotate around 200,000 samples, which are utilized to train the model. Subsequently, we use this model to infer event probabilities for the coarsely filtered headlines and filter them using a predefined threshold, achieving a 95\% accuracy rate in detecting event-related headlines.

\subsection{Event Reformulation}\label{sec:EventReformulation}

\begin{figure}[htb]
\vspace{0.0em}
\centering
\setlength{\abovecaptionskip}{0cm}    
\includegraphics[height=0.3\textheight]{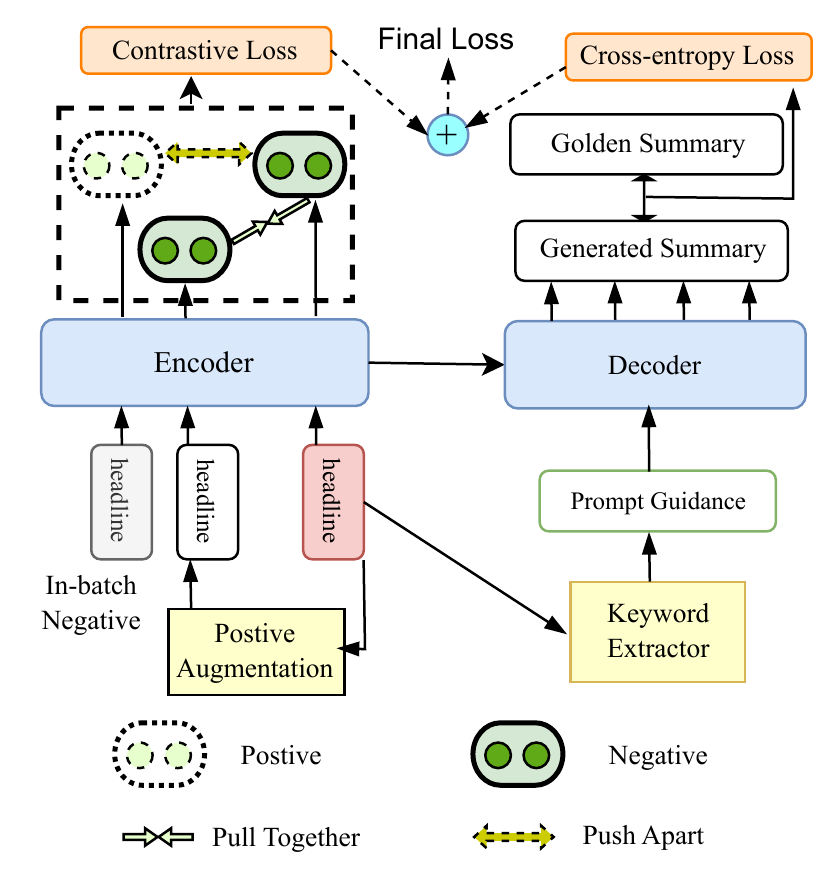}
\caption{Structure of event reformulation model.}
\label{fig:framework}
\vspace{0em}
\end{figure}

This step aims to make events qualified for query expansion by reformulating them using a generation model, addressing issues such as noise, irregular grammar, and low-frequency words. As illustrated in Figure~\ref{fig:framework}, we introduce two significant improvements to the encoder-decoder architecture-based model. Firstly, we enhance the controllability of the generation process using prompt learning. Secondly, we optimize the representation quality of headlines using contrastive learning. By simultaneously optimizing these two tasks, we can effectively refine events for query expansion.

\noindent \textbf{Prompt Guidance.}
To ensure important information is not overlooked, we leverage prompt learning technology when training a generation model. Unlike prior work, we propose adaptive keyword templates to provide guidance during sentence generation. Firstly, we use the KeyBERT model~\citep{grootendorst2020keybert} to extract the most essential nouns from the sentence. We then insert the extracted keyword into a fixed template to form a keyword template, denoted as $T$. Finally, we concatenate the headline $H$, the keyword template $T$, and the target qualified event $E$ with special tokens as ``\texttt{[CLS]}$H$ \texttt{[SEP]}\texttt{[CLS]}$T$$E$\texttt{[SEP]}''\footnote{For the BART model, the start token ID for the decoder is \texttt{[CLS]}, while for the mT5 model, it is \texttt{[PAD]}.}. In this setup, the front segment ``\texttt{[CLS]}$H$\texttt{[SEP]}'' and the latter segment ``\texttt{[CLS]}$T$$E$\texttt{[SEP]}'' serve as the inputs for the encoder and decoder, respectively. It is worth noting that $E$ is omitted during the inference phase.

\noindent \textbf{Contrastive Learning.}
Previous studies show that natural language generation tasks suffer from representation space degradation problems, which can be alleviated by contrastive learning~\citep{gao2019representation}. In our model, the embedding corresponding to the \texttt{[CLS]} token of the encoder is regarded as the headline representation and contrastive learning is performed based on it. Specifically, for each headline, we perform a position swap of its two terms to obtain a positive example headline and then use contrastive learning to pull the representations of positive headline pairs closer and push away the representations of negative headline pairs~(i.e., in-batch negative samples).

The following is a description of the contrastive learning loss calculation process within a batch.
\begin{enumerate}[label=(\alph*)]
    \item  In a batch of size 2N,  the training data consists of 2N pairs denoted by $\{(H_1, E_1), \newline (H_1^+, E_1), \cdots, (H_N, E_N), (H_N^+, E_N)\}$, where $<H_i, H_i^+>$ denote a pair of similar headlines, both of which can be paired with the event phrase $E_i$. Contrastive learning aims to pull semantically close neighbors (i.e. $(H_i, H_i^+)$) together and pushing apart non-neighbors (i.e. $(H_i, H_j^+)$, $i, j \in \{0,1,...,N\}$ and $j \neq i $).
    \item  The above 2N samples are passed through the encoder to obtain 2N embeddings that are denoted as $(e_1, e_2, .. ,e_N; e_1^+, e_2^+, ... ,e_N^+)$.
    \item  The 2N embeddings are used to compute the contrastive learning loss, which is included as part of the loss for this mini-batch. Let $\tau$ denote the temperature hyper-parameter and $\text{sim}(e_1, e_2)$ denote the cosine similarity. Then the contrastive learning loss, denoted as $L_{cl}$ is:
    \begin{equation}
        L_{cl} = - \sum\limits_{i=1}^{N} \log  \frac{e^{\text{sim}(e_i, e_i^+)/\tau}}{\sum_{j=1}^N e^{\text{sim}(e_i,e_j^+)/\tau}}\nonumber \\
    \end{equation}
 \end{enumerate}

The training data for the Event Reformulation stage consists of pairs <$headline$, $event$> that are sampled from user-click logs. Specifically, we employ two methods to construct the target event: regarding the user query as the target event and extracting the event from the headline using the LTP toolkit. To ensure that the constructed events are qualified as the target for the generation model, we further filter out pairs that do not meet our standards for loyalty, integrity, and cleanness through expert annotation.

\subsection{Semantic Retrieval}\label{sec:SemanticRetrieval}

In this step, we use a dual-tower semantic model based on contrastive learning to retrieve highly satisfying events for a given query. The work of \citet{LeeXiong2020ApproximateNN} points out that the dominance of uninformative negative samples leads to a bottleneck in the recall system, therefore, we employ a novel two-stage training paradigm.

For a given query, the positive samples are events obtained from the reformulated events of its clicked headlines. Then we use features, such as Jaccard Distance~\citep{jaccard1901etude}, BERTScore~\citep{zhang2019bertscore}, etc., to only keep relevant pairs as training samples. In the first stage, we use these positive samples with shuffled negative ones to finetune a naive dual-tower retrieval model and obtain the encoder. The weights of the model are initialized using the parameters of RoBERTa. After finetuning, we build an event vector library with more than $4$ million entries using this encoder. For a given query vector, based on Faiss~\citep{johnson2019billion}, the top-$K$ events are recalled according to the cosine similarity. Events located at the upper and lower thresholds of the threshold are regarded as hard neg samples, where the bounds are pre-defined by distribution statistics. In the second stage, we replace the randomly shuffled negative events with hard negative events and retrain the model initialized with the encoder obtained from the first stage. This results in the final retrieval model.

\subsection{Online Ranking}\label{sec:online_ranking}

Actually, selecting the optimal expansion requires considering multiple factors, such as relevance, event popularity, and timeliness. Therefore, we use the classic light-weight sorting model GBDT~\citep{friedman2001greedy}, which is compatible with the interpretation of online features. We incorporate three types of features to build the model: query-side, event-side, and interaction. Query-side features encompass query domain classification, entity recognition, word segmentation, and word weighting, among others, generated by existing online operators. Event-side features involve event found time and event popularity~(the size of the cluster to which an event belongs). The interaction features include semantic similarity, BM25~\citep{bm25-2009}, and entity matching between the query and event.

We describe the method of collecting training samples. For each query, we input the events obtained from the previous stage into the online search engine to obtain the search results pages. Then, we select the page that best satisfies the event intent of the query through expert annotation, and its corresponding event is labeled as a positive sample for the query, while the other events are labeled as negative samples. We obtain 50,000 samples, which are used to train the GBDT model for inferring the best query expansion.

\section{System Architecture}

In this section, we describe our baseline and EQE architecture in detail.

\subsection{Baseline Approach} \label{sec:BaselineApproach}

We first take a glance at our QE baseline, which is a query graph analysis framework. 
We devise a Query-Document click graph $\mathcal{G}$ based on the click propagation algorithm~\citep{jiang2016learning}. In order to prioritize time-sensitive queries, we limit our analysis to click-pairs from news websites that occurred within a $3$-day window. To mitigate the risk of irrelevant results, we integrate BM25 score to the adjacency matrix of the graph, denoted as $C$ where each entry $C_{i,j}$ is the weight of the edge between query $q_i$ and document $d_j$, specifically formulated as:

\setlength{\abovedisplayskip}{0pt}
\setlength{\belowdisplayskip}{3pt}
\begin{small}
\begin{eqnarray}
    C_{i,j} =  \begin{cases}
    \alpha \cdot \operatorname{BM25}(q_i,d_j) + 1, & \text{with edge} \\
    0, & \text{no edge}
    \end{cases}
\end{eqnarray}
\end{small}

\noindent where $\alpha$ (set to $0.2$) is a smoothing coefficient.

Representations of queries and documents are iteratively updated according to Eq.~(\ref{eq:baseline1}) and Eq.~(\ref{eq:baseline2}), respectively.

\setlength{\abovedisplayskip}{0pt}
\setlength{\belowdisplayskip}{3pt}
\begin{small}
\begin{gather}
    Q_{i}^{\left ( n \right ) } = \frac{1}{\left \|  {\textstyle \sum_{j}^{\left | Doc \right | }C_{i,j}\cdot D_{j}^{\left ( n- 1 \right ) } }  \right \|_2  } \sum_{j = 1}^{\left | Doc \right | } C_{i,j}\cdot D_{j}^{\left ( n- 1 \right ) }
    \label{eq:baseline1}\\
    D_{j}^{\left ( n \right ) } = \frac{1}{\left \|  {\textstyle \sum_{i = 1}^{\left | Query \right | }C_{i,j}\cdot Q_{i}^{\left ( n \right ) } }  \right \|_2  } \sum_{i = 1}^{\left | Query \right | } C_{i,j}\cdot Q_{i}^{\left ( n \right ) }
    \label{eq:baseline2}
\end{gather}
\end{small}

\noindent where $Q_{i}^{\left ( n \right )}$ and $D_{i}^{\left ( n \right )}$ are the representations of $q_i$ and $d_i$ at the $n$-th iteration respectively. After the $n$-th iteration, we perform clustering on $Q_{i}^{\left ( n \right )}$ to obtain the query clusters. For each cluster, we select the most frequent query as the expansion of other queries.

\subsection{EQE System}

Figure~\ref{fig:system} illustrates the EQE system, which can be divided into two parts: offline and online. The offline system sequentially processes streaming data from the internet, performing event collection, event reformulation, and Faiss indexing for fast response. These steps can be processed in parallel. In the online part, when a query arrives, a GBDT ranking model selects the top-$1$ candidate as the query expansion based on rich features. 

Furthermore, a rapidly updated caching system stores pairs of <$query$, $top$-$1$ $expansion$> to intercept requests to meet the time-consuming demands. Upon receiving a new query request, the system first seeks a pre-prepared QE in the cache. If not found, the system initiates further retrieval and ranking modules to obtain the QE. This QE is then returned to the main search system, while the <$query$, $expansion$> pair is written into the cache for any subsequent identical query requests. On the other hand, if a match is found, the cached result is immediately returned to the main search system. Concurrently, the caching system undergoes an asynchronous update in preparation for future requests. 
The implementation of an asynchronous execution pipeline does not boost the response delay of the mainstream search process. Therefore, the response time of the popular query is mainly consumed by querying the caching system. Only the stages of retrieval and ranking executed for infrequent queries lead to an increase in the response time of the search engine.

Finally, EQE covers nearly 50\% of online traffic, while the other half, such as those requiring explicit knowledge, has already been addressed by other intent understanding modules, is therefore not considered within this scope.  Online data indicates that the query expansion system elevates search latency by only 10 ms, evincing the efficacy of the proposed module.

\section{Experiments}
We conduct a series of comprehensive evaluations, both in offline and online environments, incorporating quantitative and qualitative aspects, to prove the advantages of EQE.

\subsection{End-to-end evaluation}
Firstly, we present the results of the implementation of the EQE system online, taking into account both offline and online metrics.

\noindent \textbf{Offline Evaluation.} We measure the offline performance of EQE using \textit{$Recall$@$K$} metric.
As illustrated in Eq.~(\ref{eq:recallk}), given a query $Q$, the clicked documents by users are denoted as $T=\left\{t_{1}, \ldots, t_{N}\right\}$, which are regarded as as the target.
The top-$K$ documents set recalled by the QE module is denoted as $I=\left\{i_{1}, \ldots, i_{k}\right\}$. \textit{$Recall$@$K$} is defined as: 
\begin{equation}
    {\rm \textit{Recall@K}} =\frac{\sum_{i=1}^{K} i_{i} \in T}{N}
    \label{eq:recallk}
\end{equation}

\begin{table}[t]
\small
\centering
\begin{tabular}{l|ccc}
\toprule
Methods         & Recall@100        & Recall@150    & Recall@200       \\ \midrule
Baseline        &   0.41            &   0.47        &  0.58     \\ 
EQE             &   0.58  &  0.65    & 0.74     \\
Improve            &   + 41.46\%  &  + 38.29\%     & + 27.58\%      \\ 
\bottomrule
\end{tabular}
\caption{Offline performance comparison.}
\label{tab:Offline}
\end{table}

\noindent We first collect user click-log over a certain period of time, where documents are retrieved by original queries without the influence of QE. Specifically, in our scenario, we collect news, videos, and user-generated content~(UGC). Meanwhile, we record documents retrieved by both expansions produced by EQE and the baseline approach. After 7 days of accumulation, a total of 850,000 valid online requests are collected.
As shown in Table~\ref{tab:Offline}, after evaluating \textit{$Recall$@$K$} at different thresholds, it can be seen that EQE significantly surpasses the online baseline.

\noindent \textbf{Online Evaluation.}  We construct a 30-day A/B experiment with 1\% of online traffic to gather feedback from millions of users and study the online performance of the EQE compared to the strong baseline described in Section~\ref{sec:BaselineApproach}. QEs derived from both frameworks are utilized in downstream tasks~(document retrieval and sorting).
For online evaluation, we are mainly concerned about the business metrics that impact user experience, such as $\Delta \rm{GSB}$~\citep{zou2021pre}, CTR~\citep{rangadurai2022nxtpost}, PCTR and UCTR~\citep{qin2022multi}. As shown in Table~\ref{tab:Online}, EQE outperforms the baseline and gains improvements of $6.44\%$, $6.23\%$ and $5.03\%$ on CTR, PCTR and UCTR, respectively, indicating its SOTA performance.

\begin{table}[t]
\small
\centering
\begin{tabular}{lcccc}
\toprule
~  & $\Delta\rm{GSB}$ & CTR  & PCTR   & UCTR     \\\midrule
EQE  &+12.5\%   &+6.64\%   &+6.23\%  &+5.03\%      \\ 
\bottomrule
\end{tabular} 
\caption{Online A/B test of EQE implemented.}
\label{tab:Online}
\vspace{-1.5em}
\end{table}

\subsection{Performance of Event Collection} 
We choose the intersection of ``Hot Search List'' from various platforms as our evaluation set. This decision serves two purposes: firstly, it can eliminate unfair comparisons due to platform-specific content biases; secondly, these events are highly representative in the search domain, as users consistently demonstrate in them and desire to retrieve relevant information swiftly. We employ two annotation experts to assist in the evaluation process, which involved: 1) Collecting these events from different platforms~(such as Baidu and Weibo) to find the earliest time they appeared respectively. Admittedly, since we cannot accurately determine the initial creation time of these events on other platforms, we resort to the first appearance time on the ``Hot Search List'' as an approximation; 2) Identifying the time when the first publish emerged on the Internet; 3) Recording events discovery coverage rate at several time points. 

\begin{figure}[hbtp]
\vspace{-0.5em}
    \centering
    \includegraphics[width=0.48\textwidth]{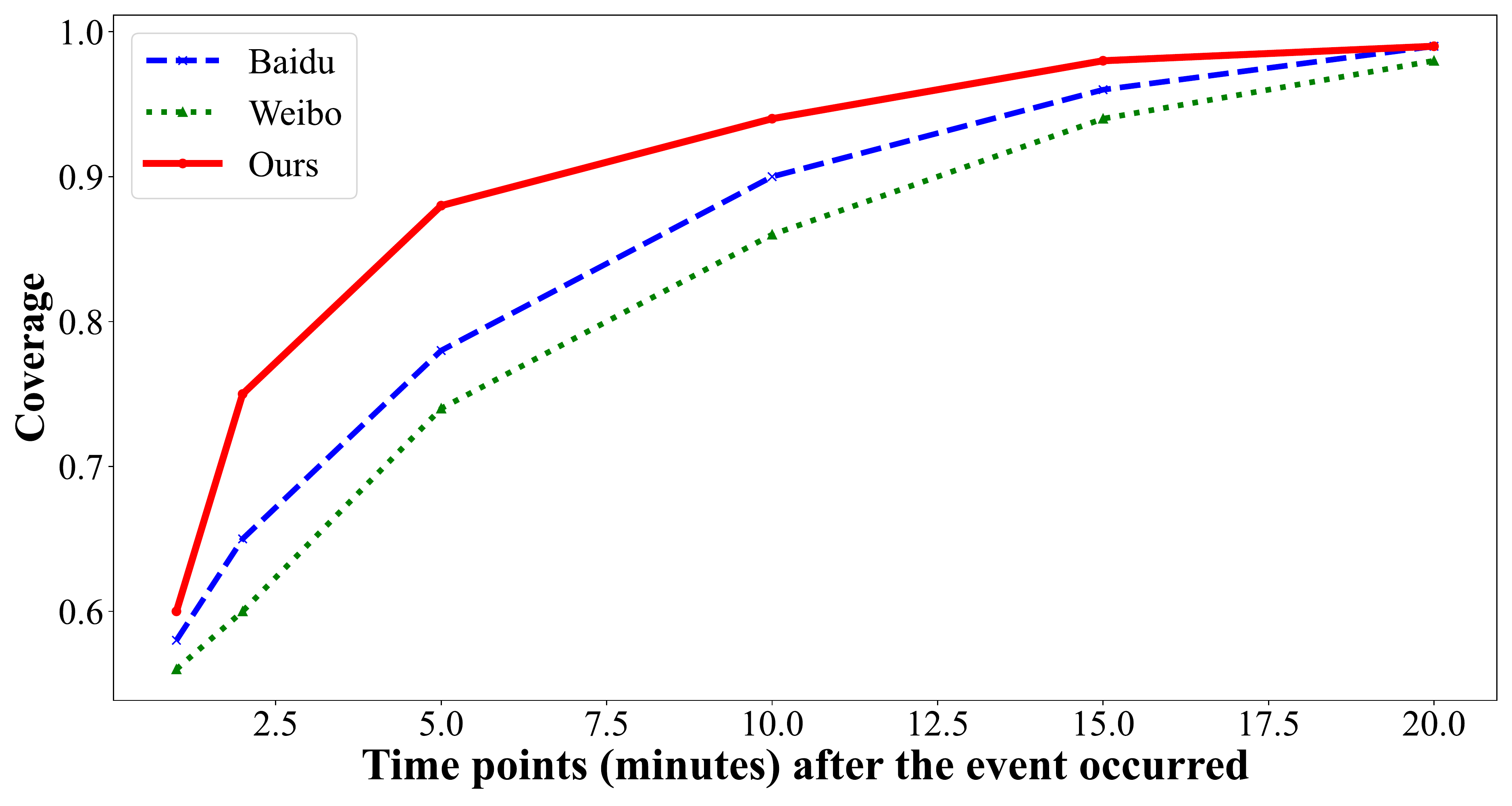}
    \caption{The x-axis represents the time points (in minutes) after the earliest occurrence of events on the internet, while the y-axis represents the coverage rate of events discovery for each system.}
    \label{fig:intime}
\vspace{-0.5em}
\end{figure}
Figure~\ref{fig:intime} illustrates the average coverage of Baidu, Weibo, and EQE at different time points after ``Hot Search List'' events occurred. The coverage of each system is recorded at time points of $1, 2, 5, 10, 15,$ and $20$ minutes. As shown, at the $5$-minute time point, EQE discovers more than $10\%$ of the events in advance compared to the other systems.

\subsection{Performance of Event Reformulation} 

We evaluate the performance of the proposed generation model by computing automated metrics. We disclose an annotated test dataset, which is called Title2EventPhrase. Production procedures and analysis of Title2EventPhrase are introduced in Appendix~\ref{sec:Title2EventPhrase_dataset} and \ref{sec:Title2EventPhrase_ana}. We conduct ablation experiments to measure the effect of the prompt and contrastive learning modules. Furthermore, to verify the universality of these two modules, we utilize BART~\citep{bart} and mT5~\citep{mt5} as the backbone networks, respectively.\urlstyle{same} \footnote{We use a popular version of the Chinese BART model available at \url{https://huggingface.co/fnlp/bart-large-chinese}, and the base version of mT5 available at \url{https://huggingface.co/google/mt5-base}. It is worth noting that the number of transformer layers in both models is consistent.} Experiments are measured with the ROUGE~\citep{lin-2004-rouge}, BLEU~\citep{papineni-etal-2002-bleu} and BERTScore metrics. As shown in Table~\ref{tab:automated}, our proposed components significantly improved the generation performance.

\begin{table}[t]
\small
\begin{tabular}{l|ccc}
\toprule
Models         & Rouge-L & BLEU & BERTScore \\ \midrule
BART~(vanilla) &   0.8391    &   0.7692    &  0.9266 \\ 
BART + CL    &    0.8406   &   0.7724  & 0.9278 \\ 
BART + PG    &  0.8458 & 0.7777 & 0.9294 \\ 
BART + CL + PG    & \textbf{0.8480}  & \textbf{0.7822}  &  \textbf{0.9312} \\ \midrule
mT5~(vanilla)  &   0.8453    &   0.7781    &  0.9297   \\  
mT5 + CL  &  0.8489     &   0.7833   &   0.9315    \\ 
mT5 + PG  &  0.8511    & 0.7857   &  0.9322 \\ 
mT5 + CL + PG   &  \textbf{0.8533}      &  \textbf{0.7897}  & \textbf{0.9336} \\ \bottomrule
\end{tabular}
\caption{Automated evaluation of ablation experiments. CL and PG are abbreviations for contrastive learning and prompt guidance, respectively.}
\label{tab:automated}
\end{table}

\subsection{Performance of Semantic Retrieval} 
In this section, we evaluate the effectiveness of our proposed semantic retrieval model against the baseline by employing three definitive performance metrics: standard \text{Recall@K}~\citep{jegou2010product}, \text{MRR@K}~\citep{Craswell2009} and \text{AUC}~\citep{FAWCETT2006861}.

We construct a testing dataset with a similar method to obtain <$query$, $event$> pairs as the training dataset mentioned in Section~\ref{sec:SemanticRetrieval}, with relevance labels annotated by experts. We sample them at different time periods to ensure training and testing datasets have the same distribution but are non-overlapping. Table~\ref{tab:semantic} shows the advantages of our training scheme over other baseline models. In addition, we also visualize the results of a two-dimensional t-SNE~\citep{van2008visualizing} graph on the embedding of 100,000 queries, further demonstrating the effectiveness of our proposed method in addressing the problem of representation degradation. For more details, please refer to Appendix~\ref{sec:ContrastiveLearningRepresentations_ana} and Figure~\ref{fig:tsne}.

\begin{table}[t]
\small
\begin{tabular}{l|ccc}
\toprule
Models         & Recall@10 & MRR@10 & AUC \\ \midrule
RoBERTa~(vanilla) &   0.74    &   0.43    &  0.80 \\ 
RoBERTa + CL    &   0.75  &  0.45  & 0.82 \\ 
RoBERTa + CL + 2T    &  \textbf{0.80}  &  \textbf{0.51}  & \textbf{0.87} \\ \bottomrule
\end{tabular} 
\caption{Evaluation of semantic models. CL and 2T are abbreviations for contrastive learning and two-stage training with hard neg samples, respectively.}
\label{tab:semantic}
\vspace{-1em}
\end{table}

\section{Related Work}
Query understanding~(\textbf{QU}) is a fundamental task of information retrieval~(\textbf{IR}), which aims to help reformulate query, predict query intent, and ultimately improve the document relevance modeling~\citep{zhang2020query}.
As an essential method for QU, query expansion (\textbf{QE}) involves the addition of relevant terms or specific information to a query to clarify intention and enhance retrieval performance ~\citep{rosin2021event}. In recent years, most QE methods have been based on word embedding techniques~\citep{srinivasan2022quill, padaki2020rethinking, azad2019new, kuzi2016query, zamani2016embedding}, which select semantically related terms as expansions.
Usually, word embeddings are learned in two ways, one is based on the semantic vector of terms and the other is based on retrieval relevance~\citep{diaz2016query, zamani2017relevance}. Meanwhile, external data sources, such as Wikipedia and WordNet, have also been utilized for QE ~\citep{azad2019new}. \looseness=-1

However, these conventional QE methods mainly rely on mining search logs or pre-built expansion libraries, which leads to slow update rates in time-sensitive scenarios. On the other hand, new occurring events in real time can meet the timeliness requirements well, and mining QE from them is a promising research direction.
Recently \citet{deng-etal-2022-title2event} construct a large-scale dataset aiming at extracting event arguments, like \textit{subject}, \textit{predicate} and \textit{object}, from news headlines. 
However, structured outputs from extractive models~\cite{lu-etal-2022-unified,gao-etal-2022-mask} might not be fully utilized by the retrieval and ranking modules.  
We thus turn to generative models for event reformation. 
Normalized events serve as crucial signals for time-sensitive query expansion, which makes the largest contribution to our work.  

\section{Conclusion}

This paper presents our solution for large-scale event-centric query expansion, called EQE, which is able to efficiently capture query intent for ongoing events and help our search engine to retrieve more event-related results. Advanced techniques are involved in each stage of EQE to improve performance. Offline experiments and online A/B tests verify the effectiveness of EQE. We have deployed the system in Tencent QQ Browser Search to serve hundreds of millions of users. Meanwhile, we share the design and deployment scheme.

\bibliography{anthology,custom}

\begin{thebibliography}{42}
\expandafter\ifx\csname natexlab\endcsname\relax\def\natexlab#1{#1}\fi

\bibitem[{Azad and Deepak(2019)}]{azad2019new}
Hiteshwar~Kumar Azad and Akshay Deepak. 2019.
\newblock A new approach for query expansion using wikipedia and wordnet.
\newblock \emph{Information sciences}, 492:147--163.

\bibitem[{Broder et~al.(2007)Broder, Fontoura, Gabrilovich, Joshi, Josifovski,
  and Zhang}]{broder2007robust}
Andrei~Z Broder, Marcus Fontoura, Evgeniy Gabrilovich, Amruta Joshi, Vanja
  Josifovski, and Tong Zhang. 2007.
\newblock Robust classification of rare queries using web knowledge.
\newblock In \emph{Proceedings of the 30th annual international ACM SIGIR
  conference on Research and development in information retrieval}, pages
  231--238.

\bibitem[{Caruccio et~al.(2015)Caruccio, Deufemia, and
  Polese}]{caruccio2015understanding}
Loredana Caruccio, Vincenzo Deufemia, and Giuseppe Polese. 2015.
\newblock Understanding user intent on the web through interaction mining.
\newblock \emph{Journal of Visual Languages \& Computing}, 31:230--236.

\bibitem[{Che et~al.(2021)Che, Feng, Qin, and Liu}]{che-etal-2021-n}
Wanxiang Che, Yunlong Feng, Libo Qin, and Ting Liu. 2021.
\newblock \href {https://doi.org/10.18653/v1/2021.emnlp-demo.6} {N-{LTP}: An
  open-source neural language technology platform for {C}hinese}.
\newblock In \emph{Proceedings of the 2021 Conference on Empirical Methods in
  Natural Language Processing: System Demonstrations}, pages 42--49, Online and
  Punta Cana, Dominican Republic. Association for Computational Linguistics.

\bibitem[{Craswell(2009)}]{Craswell2009}
Nick Craswell. 2009.
\newblock \href {https://doi.org/10.1007/978-0-387-39940-9_488} {\emph{Mean
  Reciprocal Rank}}, pages 1703--1703. Springer US, Boston, MA.

\bibitem[{Deng et~al.(2022)Deng, Zhang, Zhang, Ying, Yu, Gao, Wang, Bai, Yang,
  Ma, Chen, and Zhou}]{deng-etal-2022-title2event}
Haolin Deng, Yanan Zhang, Yangfan Zhang, Wangyang Ying, Changlong Yu, Jun Gao,
  Wei Wang, Xiaoling Bai, Nan Yang, Jin Ma, Xiang Chen, and Tianhua Zhou. 2022.
\newblock \href {https://aclanthology.org/2022.emnlp-main.437}
  {{T}itle2{E}vent: Benchmarking open event extraction with a large-scale
  {C}hinese title dataset}.
\newblock In \emph{Proceedings of the 2022 Conference on Empirical Methods in
  Natural Language Processing}, pages 6511--6524, Abu Dhabi, United Arab
  Emirates. Association for Computational Linguistics.

\bibitem[{Diaz et~al.(2016)Diaz, Mitra, and Craswell}]{diaz2016query}
Fernando Diaz, Bhaskar Mitra, and Nick Craswell. 2016.
\newblock \href {https://doi.org/10.18653/v1/P16-1035} {Query expansion with
  locally-trained word embeddings}.
\newblock In \emph{Proceedings of the 54th Annual Meeting of the Association
  for Computational Linguistics (Volume 1: Long Papers)}, pages 367--377,
  Berlin, Germany. Association for Computational Linguistics.

\bibitem[{Fawcett(2006)}]{FAWCETT2006861}
Tom Fawcett. 2006.
\newblock \href {https://doi.org/https://doi.org/10.1016/j.patrec.2005.10.010}
  {An introduction to roc analysis}.
\newblock \emph{Pattern Recognition Letters}, 27(8):861--874.
\newblock ROC Analysis in Pattern Recognition.

\bibitem[{Friedman(2001)}]{friedman2001greedy}
Jerome~H Friedman. 2001.
\newblock Greedy function approximation: a gradient boosting machine.
\newblock In \emph{Annals of statistics}, pages 1189--1232. JSTOR.

\bibitem[{Gao et~al.(2019)Gao, He, Tan, Qin, Wang, and
  Liu}]{gao2019representation}
Jun Gao, Di~He, Xu~Tan, Tao Qin, Liwei Wang, and Tie{-}Yan Liu. 2019.
\newblock \href {https://openreview.net/forum?id=SkEYojRqtm} {Representation
  degeneration problem in training natural language generation models}.
\newblock In \emph{7th International Conference on Learning Representations,
  {ICLR} 2019, New Orleans, LA, USA, May 6-9, 2019}. OpenReview.net.

\bibitem[{Gao et~al.(2022)Gao, Yu, Wang, Zhao, and Xu}]{gao-etal-2022-mask}
Jun Gao, Changlong Yu, Wei Wang, Huan Zhao, and Ruifeng Xu. 2022.
\newblock \href {https://aclanthology.org/2022.findings-emnlp.332}
  {Mask-then-fill: A flexible and effective data augmentation framework for
  event extraction}.
\newblock In \emph{Findings of the Association for Computational Linguistics:
  EMNLP 2022}, pages 4537--4544, Abu Dhabi, United Arab Emirates. Association
  for Computational Linguistics.

\bibitem[{Gao et~al.(2021)Gao, Yao, and Chen}]{gao2021simcse}
Tianyu Gao, Xingcheng Yao, and Danqi Chen. 2021.
\newblock \href {https://doi.org/10.18653/v1/2021.emnlp-main.552} {{S}im{CSE}:
  Simple contrastive learning of sentence embeddings}.
\newblock In \emph{Proceedings of the 2021 Conference on Empirical Methods in
  Natural Language Processing}, pages 6894--6910, Online and Punta Cana,
  Dominican Republic. Association for Computational Linguistics.

\bibitem[{Gillick et~al.(2019)Gillick, Kulkarni, Lansing, Presta, Baldridge,
  Ie, and Garcia-Olano}]{DanielGillick2019LearningDR}
Daniel Gillick, Sayali Kulkarni, Larry Lansing, Alessandro Presta, Jason
  Baldridge, Eugene Ie, and Diego Garcia-Olano. 2019.
\newblock \href {https://doi.org/10.18653/v1/K19-1049} {Learning dense
  representations for entity retrieval}.
\newblock In \emph{Proceedings of the 23rd Conference on Computational Natural
  Language Learning (CoNLL)}, pages 528--537, Hong Kong, China. Association for
  Computational Linguistics.

\bibitem[{Grootendorst(2020)}]{grootendorst2020keybert}
Maarten Grootendorst. 2020.
\newblock \href {https://doi.org/10.5281/zenodo.4461265} {Keybert: Minimal
  keyword extraction with bert.}

\bibitem[{Jaccard(1901)}]{jaccard1901etude}
Paul Jaccard. 1901.
\newblock Étude comparative de la distribution florale dans une portion des
  alpes et des jura.
\newblock \emph{Bulletin del la Société Vaudoise des Sciences Naturelles},
  37:547--579.

\bibitem[{Jansen et~al.(2007)Jansen, Booth, and Spink}]{jansen2007determining}
Bernard~J. Jansen, Danielle~L. Booth, and Amanda Spink. 2007.
\newblock \href {https://doi.org/10.1145/1242572.1242739} {Determining the user
  intent of web search engine queries}.
\newblock In \emph{Proceedings of the 16th International Conference on World
  Wide Web, {WWW} 2007, Banff, Alberta, Canada, May 8-12, 2007}, pages
  1149--1150. {ACM}.

\bibitem[{Jegou et~al.(2010)Jegou, Douze, and Schmid}]{jegou2010product}
Herve Jegou, Matthijs Douze, and Cordelia Schmid. 2010.
\newblock Product quantization for nearest neighbor search.
\newblock \emph{IEEE transactions on pattern analysis and machine
  intelligence}, 33(1):117--128.

\bibitem[{Jiang et~al.(2016)Jiang, Hu, Kang, Jr., Yin, Chang, and
  Zhai}]{jiang2016learning}
Shan Jiang, Yuening Hu, Changsung Kang, Tim~Daly Jr., Dawei Yin, Yi~Chang, and
  ChengXiang Zhai. 2016.
\newblock \href {https://doi.org/10.1145/2911451.2911531} {Learning query and
  document relevance from a web-scale click graph}.
\newblock In \emph{Proceedings of the 39th International {ACM} {SIGIR}
  conference on Research and Development in Information Retrieval, {SIGIR}
  2016, Pisa, Italy, July 17-21, 2016}, pages 185--194. {ACM}.

\bibitem[{Johnson et~al.(2019)Johnson, Douze, and
  J{\'e}gou}]{johnson2019billion}
Jeff Johnson, Matthijs Douze, and Herv{\'e} J{\'e}gou. 2019.
\newblock Billion-scale similarity search with {GPUs}.
\newblock \emph{IEEE Transactions on Big Data}, 7(3):535--547.

\bibitem[{Kuzi et~al.(2016)Kuzi, Shtok, and Kurland}]{kuzi2016query}
Saar Kuzi, Anna Shtok, and Oren Kurland. 2016.
\newblock \href {https://doi.org/10.1145/2983323.2983876} {Query expansion
  using word embeddings}.
\newblock In \emph{Proceedings of the 25th {ACM} International Conference on
  Information and Knowledge Management, {CIKM} 2016, Indianapolis, IN, USA,
  October 24-28, 2016}, pages 1929--1932. {ACM}.

\bibitem[{Lewis et~al.(2020)Lewis, Liu, Goyal, Ghazvininejad, Mohamed, Levy,
  Stoyanov, and Zettlemoyer}]{bart}
Mike Lewis, Yinhan Liu, Naman Goyal, Marjan Ghazvininejad, Abdelrahman Mohamed,
  Omer Levy, Veselin Stoyanov, and Luke Zettlemoyer. 2020.
\newblock \href {https://doi.org/10.18653/v1/2020.acl-main.703} {{BART}:
  Denoising sequence-to-sequence pre-training for natural language generation,
  translation, and comprehension}.
\newblock In \emph{Proceedings of the 58th Annual Meeting of the Association
  for Computational Linguistics}, pages 7871--7880, Online. Association for
  Computational Linguistics.

\bibitem[{Lin(2004)}]{lin-2004-rouge}
Chin-Yew Lin. 2004.
\newblock \href {https://aclanthology.org/W04-1013} {{ROUGE}: A package for
  automatic evaluation of summaries}.
\newblock In \emph{Text Summarization Branches Out}, pages 74--81, Barcelona,
  Spain. Association for Computational Linguistics.

\bibitem[{Liu et~al.(2019)Liu, Ott, Goyal, Du, Joshi, Chen, Levy, Lewis,
  Zettlemoyer, and Stoyanov}]{YinhanLiu2019RoBERTaAR}
Yinhan Liu, Myle Ott, Naman Goyal, Jingfei Du, Mandar Joshi, Danqi Chen, Omer
  Levy, Michael Lewis, Luke Zettlemoyer, and Veselin Stoyanov. 2019.
\newblock Roberta: A robustly optimized bert pretraining approach.
\newblock \emph{Cornell University - arXiv}.

\bibitem[{Lu et~al.(2022)Lu, Liu, Dai, Xiao, Lin, Han, Sun, and
  Wu}]{lu-etal-2022-unified}
Yaojie Lu, Qing Liu, Dai Dai, Xinyan Xiao, Hongyu Lin, Xianpei Han, Le~Sun, and
  Hua Wu. 2022.
\newblock \href {https://doi.org/10.18653/v1/2022.acl-long.395} {Unified
  structure generation for universal information extraction}.
\newblock In \emph{Proceedings of the 60th Annual Meeting of the Association
  for Computational Linguistics (Volume 1: Long Papers)}, pages 5755--5772,
  Dublin, Ireland. Association for Computational Linguistics.

\bibitem[{Nogueira et~al.(2019)Nogueira, Lin, and
  Epistemic}]{nogueira2019doc2query}
Rodrigo Nogueira, Jimmy Lin, and AI~Epistemic. 2019.
\newblock From doc2query to doctttttquery.
\newblock \emph{Online preprint}, 6.

\bibitem[{Padaki et~al.(2020)Padaki, Dai, and Callan}]{padaki2020rethinking}
Ramith Padaki, Zhuyun Dai, and Jamie Callan. 2020.
\newblock Rethinking query expansion for bert reranking.
\newblock In \emph{Advances in Information Retrieval: 42nd European Conference
  on IR Research, ECIR 2020, Lisbon, Portugal, April 14--17, 2020, Proceedings,
  Part II 42}, pages 297--304. Springer.

\bibitem[{Papineni et~al.(2002)Papineni, Roukos, Ward, and
  Zhu}]{papineni-etal-2002-bleu}
Kishore Papineni, Salim Roukos, Todd Ward, and Wei-Jing Zhu. 2002.
\newblock \href {https://doi.org/10.3115/1073083.1073135} {{B}leu: a method for
  automatic evaluation of machine translation}.
\newblock In \emph{Proceedings of the 40th Annual Meeting of the Association
  for Computational Linguistics}, pages 311--318, Philadelphia, Pennsylvania,
  USA. Association for Computational Linguistics.

\bibitem[{Qin et~al.(2022)Qin, Wang, Ma, and Zhang}]{qin2022multi}
Yuqi Qin, Pengfei Wang, Biyu Ma, and Zhe Zhang. 2022.
\newblock A multi-interest evolution story: Applying psychology in query-based
  recommendation for inferring customer intention.
\newblock In \emph{Proceedings of the 31st ACM International Conference on
  Information \& Knowledge Management}, pages 1655--1665.

\bibitem[{Rangadurai et~al.(2022)Rangadurai, Liu, Malreddy, Liu, Maheshwari,
  Sangale, and Borisyuk}]{rangadurai2022nxtpost}
Kaushik Rangadurai, Yiqun Liu, Siddarth Malreddy, Xiaoyi Liu, Piyush
  Maheshwari, Vishwanath Sangale, and Fedor Borisyuk. 2022.
\newblock Nxtpost: User to post recommendations in facebook groups.
\newblock In \emph{Proceedings of the 28th ACM SIGKDD Conference on Knowledge
  Discovery and Data Mining}, pages 3792--3800.

\bibitem[{Robertson and Zaragoza(2009)}]{bm25-2009}
Stephen Robertson and Hugo Zaragoza. 2009.
\newblock \href {https://doi.org/10.1561/1500000019} {The probabilistic
  relevance framework: Bm25 and beyond}.
\newblock \emph{Foundations and Trends in Information Retrieval}, 3:333--389.

\bibitem[{Rosin et~al.(2021)Rosin, Guy, and Radinsky}]{rosin2021event}
Guy~D Rosin, Ido Guy, and Kira Radinsky. 2021.
\newblock Event-driven query expansion.
\newblock In \emph{Proceedings of the 14th ACM International Conference on Web
  Search and Data Mining}, pages 391--399.

\bibitem[{Srinivasan et~al.(2022)Srinivasan, Raman, Samanta, Liao, Bertelli,
  and Bendersky}]{srinivasan2022quill}
Krishna Srinivasan, Karthik Raman, Anupam Samanta, Lingrui Liao, Luca Bertelli,
  and Mike Bendersky. 2022.
\newblock \href {https://arxiv.org/abs/2210.15718} {Quill: Query intent with
  large language models using retrieval augmentation and multi-stage
  distillation}.
\newblock \emph{ArXiv preprint}, abs/2210.15718.

\bibitem[{Sun et~al.(2022)Sun, Lu, Ma, Liu, and Guo}]{Sun2022}
Zhongkai Sun, Sixing Lu, Chengyuan Ma, Xiaohu Liu, and Edward Guo. 2022.
\newblock \href
  {https://www.amazon.science/publications/query-expansion-and-entity-weighting-for-query-reformulation-retrieval-in-voice-assistant-systems}
  {Query expansion and entity weighting for query reformulation retrieval in
  voice assistant systems}.
\newblock In \emph{WSDM 2022}.

\bibitem[{Van~der Maaten and Hinton(2008)}]{van2008visualizing}
Laurens Van~der Maaten and Geoffrey Hinton. 2008.
\newblock Visualizing data using t-sne.
\newblock \emph{Journal of machine learning research}, 9(11).

\bibitem[{Xiong et~al.(2021)Xiong, Xiong, Li, Tang, Liu, Bennett, Ahmed, and
  Overwijk}]{LeeXiong2020ApproximateNN}
Lee Xiong, Chenyan Xiong, Ye~Li, Kwok{-}Fung Tang, Jialin Liu, Paul~N. Bennett,
  Junaid Ahmed, and Arnold Overwijk. 2021.
\newblock \href {https://openreview.net/forum?id=zeFrfgyZln} {Approximate
  nearest neighbor negative contrastive learning for dense text retrieval}.
\newblock In \emph{9th International Conference on Learning Representations,
  {ICLR} 2021, Virtual Event, Austria, May 3-7, 2021}. OpenReview.net.

\bibitem[{Xue et~al.(2021)Xue, Constant, Roberts, Kale, Al-Rfou, Siddhant,
  Barua, and Raffel}]{mt5}
Linting Xue, Noah Constant, Adam Roberts, Mihir Kale, Rami Al-Rfou, Aditya
  Siddhant, Aditya Barua, and Colin Raffel. 2021.
\newblock \href {https://doi.org/10.18653/v1/2021.naacl-main.41} {m{T}5: A
  massively multilingual pre-trained text-to-text transformer}.
\newblock In \emph{Proceedings of the 2021 Conference of the North American
  Chapter of the Association for Computational Linguistics: Human Language
  Technologies}, pages 483--498, Online. Association for Computational
  Linguistics.

\bibitem[{Zamani and Croft(2016)}]{zamani2016embedding}
Hamed Zamani and W~Bruce Croft. 2016.
\newblock Embedding-based query language models.
\newblock In \emph{Proceedings of the 2016 ACM international conference on the
  theory of information retrieval}, pages 147--156.

\bibitem[{Zamani and Croft(2017)}]{zamani2017relevance}
Hamed Zamani and W.~Bruce Croft. 2017.
\newblock \href {https://doi.org/10.1145/3077136.3080831} {Relevance-based word
  embedding}.
\newblock In \emph{Proceedings of the 40th International {ACM} {SIGIR}
  Conference on Research and Development in Information Retrieval, Shinjuku,
  Tokyo, Japan, August 7-11, 2017}, pages 505--514. {ACM}.

\bibitem[{Zamora et~al.(2014)Zamora, Mendoza, and Allende}]{zamora2014query}
Juan Zamora, Marcelo Mendoza, and H{\'e}ctor Allende. 2014.
\newblock Query intent detection based on query log mining.
\newblock \emph{J. Web Eng.}, 13(1\&2):24--52.

\bibitem[{Zhang et~al.(2020{\natexlab{a}})Zhang, Guo, Fan, Lan, and
  Cheng}]{zhang2020query}
Ruqing Zhang, Jiafeng Guo, Yixing Fan, Yanyan Lan, and Xueqi Cheng.
  2020{\natexlab{a}}.
\newblock \href {https://doi.org/10.1145/3340531.3411999} {Query understanding
  via intent description generation}.
\newblock In \emph{{CIKM} '20: The 29th {ACM} International Conference on
  Information and Knowledge Management, Virtual Event, Ireland, October 19-23,
  2020}, pages 1823--1832. {ACM}.

\bibitem[{Zhang et~al.(2020{\natexlab{b}})Zhang, Kishore, Wu, Weinberger, and
  Artzi}]{zhang2019bertscore}
Tianyi Zhang, Varsha Kishore, Felix Wu, Kilian~Q. Weinberger, and Yoav Artzi.
  2020{\natexlab{b}}.
\newblock \href {https://openreview.net/forum?id=SkeHuCVFDr} {Bertscore:
  Evaluating text generation with {BERT}}.
\newblock In \emph{8th International Conference on Learning Representations,
  {ICLR} 2020, Addis Ababa, Ethiopia, April 26-30, 2020}. OpenReview.net.

\bibitem[{Zou et~al.(2021)Zou, Zhang, Cai, Ma, Cheng, Wang, Shi, Cheng, and
  Yin}]{zou2021pre}
Lixin Zou, Shengqiang Zhang, Hengyi Cai, Dehong Ma, Suqi Cheng, Shuaiqiang
  Wang, Daiting Shi, Zhicong Cheng, and Dawei Yin. 2021.
\newblock Pre-trained language model based ranking in baidu search.
\newblock In \emph{Proceedings of the 27th ACM SIGKDD Conference on Knowledge
  Discovery \& Data Mining}, pages 4014--4022.

\end{thebibliography}
\bibliographystyle{acl_natbib}

\clearpage
\appendix
\begin{figure*}[h]
\vspace{-0.5em}
    \centering
    \setlength{\abovecaptionskip}{0.cm}
    \includegraphics[width=1\textwidth]{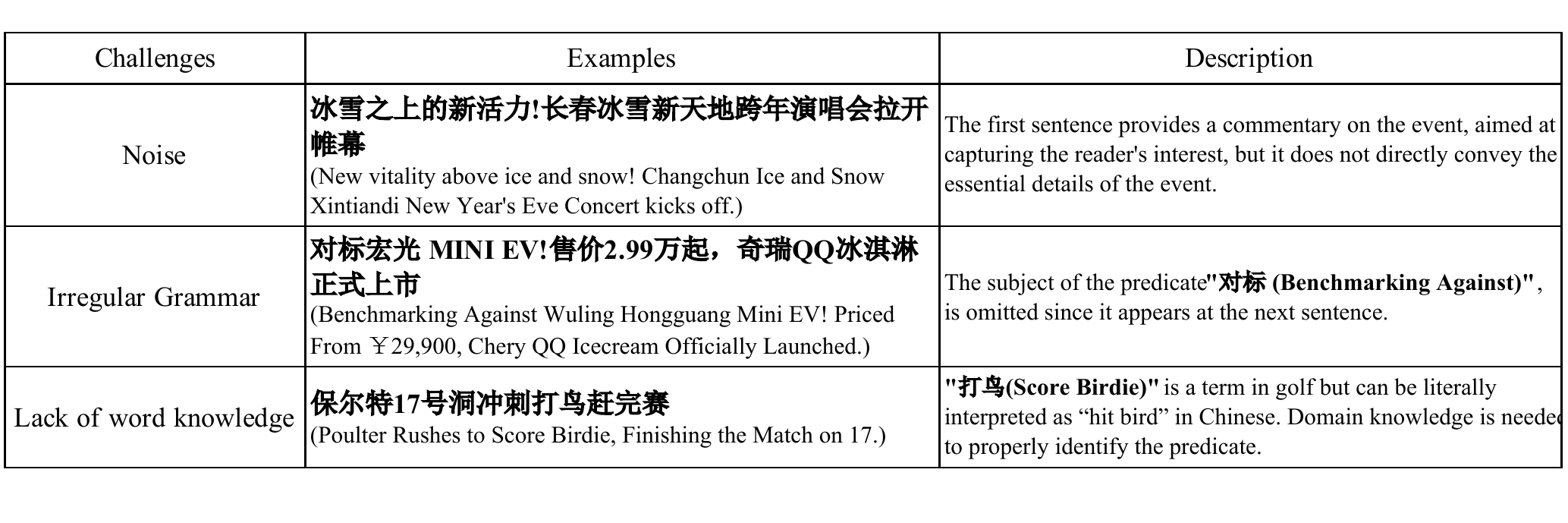}
    \caption{Challenges and examples of news headlines.}
    \label{fig:examples}
\vspace{-0.5em}
\end{figure*}
\section{Characteristics of News Headlines }\label{sec:CharacteristicsCET}
News headlines, designed to catch the reader's attention and highlight the editor's perspective, may contain redundant information beyond the event itself. In addition, they often use irregular grammar for the sake of memorability, and some words require extensive knowledge to comprehend. We provided several examples in Figure~\ref{fig:examples}, which clearly demonstrate the necessity of using generation models to reformulate them before using them as query expansion.



\section{Title2EventPhrase Construction}\label{sec:Title2EventPhrase_dataset}

\textbf{Data Collection.}
We collect a broad range of Chinese web pages from January to March 2022, using web crawler logs from Tencent QQ Browser Search, and choose a reliable business tool to identify web pages that mention events~(primarily from news websites).
Following this, we extract the titles of the chosen web pages and automatically label them with our predefined topics. Any titles that contain toxic content are removed. To ensure a diverse range of events, we conduct data sampling every ten days during the crawling period. This reduces the frequency of events that belong to the most commonly occurring topics, resulting in a more balanced distribution of topics. In total, we collected over 100,000 instances.
\noindent \\
\textbf{Annotation Standard.}
We summarize some essential parts of our annotation criteria. 
Our goal is to obtain clear and concise event descriptions from the titles and to extract the most chief~(core) events from titles that contain multiple events. To achieve this, we have outlined some important specifications below:

1) Our definition of an event is a real-world behavior or change in state. It is worth noting that statements like policy notices or subjective opinions are not considered events. Furthermore, if a title is unclear or contains multiple unrelated events~(e.g. news roundup), it should be flagged as ``invalid'' by annotators.

2) We have specified some rules to clarify the labeling of event phrases: 
\textbf{a)} definite~(non-interrogative), fluent~(good readability) description of the event in the title; 
\textbf{b)} consistent with the fact described in the title; 
\textbf{c)} if there is a progressive relationship between multiple events in the title, they need to be included to ensure the integrity of the information; 
\textbf{d)} quantifiers, gerunds, and complements need to be removed if they do not affect the understanding of the event, otherwise should be retained.
\noindent \\
\textbf{Crowdsourced Annotation.} We cooperate with crowdsourcing companies to hire human annotators. After multi-rounds of training in three weeks, 27 annotators are selected. We pay them ￥0.3 per instance. Meanwhile, four experts participated in two rounds of annotation checking for quality control. For each instance, a human annotator is asked to write the core event phase independently. To reduce the annotation difficulty, we provide a reference output along with the raw title. In the beginning, the reference output is mined from the query in click-log. After 10,000 labeled instances are collected, we train a better BART model for the rest of the annotation process. Also, we allow the annotators to refer to search engines to acquire domain knowledge. The crowd-sourced annotation is conducted in batches with two rounds of quality checking before being integrated into the final version of our dataset. 
\noindent \\
\textbf{Two rounds Checking.} Each time the crowd-sourced annotation of a batch is completed, it is sent to four experts to check whether they meet the requirements of our annotation standard. Instances that do not pass the quality check will be sent back for revision, attached with the specific reasons for rejection. This process repeats until the acceptance rate reaches 90\%. Then, the current batch is sent to the authors for dual-check. The authors will randomly check 30\% of the instances and send unqualified instances back to the experts along with the reasons for rejection. Slight adjustments on annotation standards also take place in this phase. This process repeats until the acceptance rate reaches 95\%.

\section{Title2EventPhrase Analysis}\label{sec:Title2EventPhrase_ana}

\noindent \textbf{Overview.} Table~\ref{tab:overview} shows the overview of the Title2EventPhrase dataset, including data size, topic numbers, and the average length of titles and events.

\begin{table}[H]
    \centering
    \begin{tabular}[width=\linewidth]{lc}
    \toprule
    Attribute   &   Value   \\ \midrule
    Data Size   &  41351  \\ 
    Number of Topics  &  25 \\
    Avg. Len. of Title & 25.85 \\
    Avg. Len. of Event & 16.68 \\
    \bottomrule
    \end{tabular}
    \caption{The overall statistics of Title2EventPhrase.}
    \label{tab:overview}
\end{table}

\begin{figure*}[hbtp]
\vspace{1.5em}
    \centering
    \setlength{\abovecaptionskip}{0.cm}
    \includegraphics[width=1\textwidth]{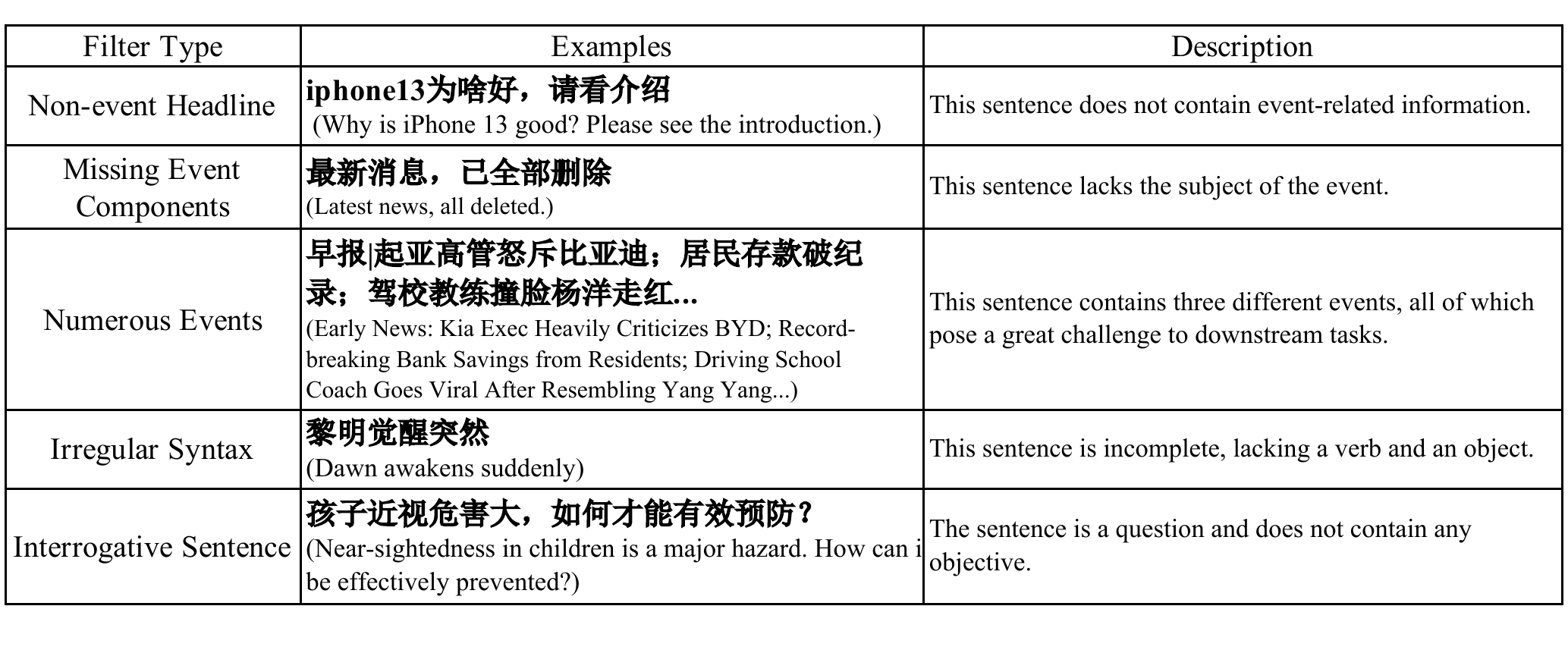}
    \caption{Filter rules in coarse filter stage.}
    \label{tab:filter_example}
\vspace{-0.5em}
\end{figure*}

\noindent \textbf{Topic Distribution.} Table~\ref{tab:topic_dis} lists 25 topics with their respective numbers and proportions. The distribution of topics in the dataset is obviously long-tailed. The largest number of topics is Society, whose proportion exceeds $31\%$, while 11 topics account for less than $1\%$.

\begin{table}
    \small
    \centering
    \begin{tabular}{lcc} 
    \toprule
    Topic & Count & Proportion \\
    \midrule
    社会~(Society) & 12985 & 31.40\% \\
    财经~(Finance) & 5877 & 14.21\% \\
    体育~(Sports) & 4504 & 10.89\% \\
    时事~(Current Events) & 4078 & 9.86\% \\
    科技~(Technology) & 2698 & 6.52\% \\
    娱乐~(Entertainment) & 1903 & 4.60\% \\
    教育~(Education) & 1415 & 3.42\% \\
    天气~(Weather) & 1307 & 3.16\% \\
    汽车~(Cars) & 1255 & 3.03\% \\
    军事~(Military) & 738 & 1.78\% \\
    房产~(Real Estate) & 614 & 1.48\% \\
    旅游~(Travel) & 597 & 1.44\% \\
    三农~(Agriculture) & 546 & 1.32\% \\
    文化~(Culture) & 435 & 1.05\% \\
    游戏~(Games) & 365 & 0.88\% \\
    综艺~(Variety Shows) & 363 & 0.88\% \\
    电影~(Movies) & 324 & 0.78\% \\
    健康~(Health) & 314 & 0.76\% \\
    电视剧~(TV Series) & 210 & 0.51\% \\
    历史~(History) & 202 & 0.49\% \\
    科学~(Science) & 150 & 0.36\% \\
    音乐~(Music) & 150 & 0.36\% \\
    生活~(Life) & 116 & 0.28\% \\
    美食~(Food) & 103 & 0.25\% \\
    情感~(Sentiment) & 102 & 0.25\% \\
    \midrule
    Total & 41351 &  100\% \\
    \bottomrule
    \end{tabular}
    \caption{The topics in Title2EventPhrase with their numbers and proportions of instances.}
    \label{tab:topic_dis}
\end{table}

\noindent \textbf{Challenge Distribution.} In this section, We analyze the scale of observed challenges described in Figure~\ref{fig:examples}. We randomly select 1,000 headlines and manually annotate which challenge type it belongs to. The annotation result shows that $27\%$ of sampled headlines have redundant expressions, $12\%$ of them suffer from irregular grammar issues, and $11\%$ of them require domain knowledge for sentence understanding.

\section{Event Representation Analysis}\label{sec:ContrastiveLearningRepresentations_ana}

As mentioned in Section~\ref{sec:SemanticRetrieval}, our baseline model suffers from the issue of representation space
degradation, leading to poor generation and retrieval performance.

\begin{figure}[h]
    \centering
    \includegraphics[width=0.48\textwidth]{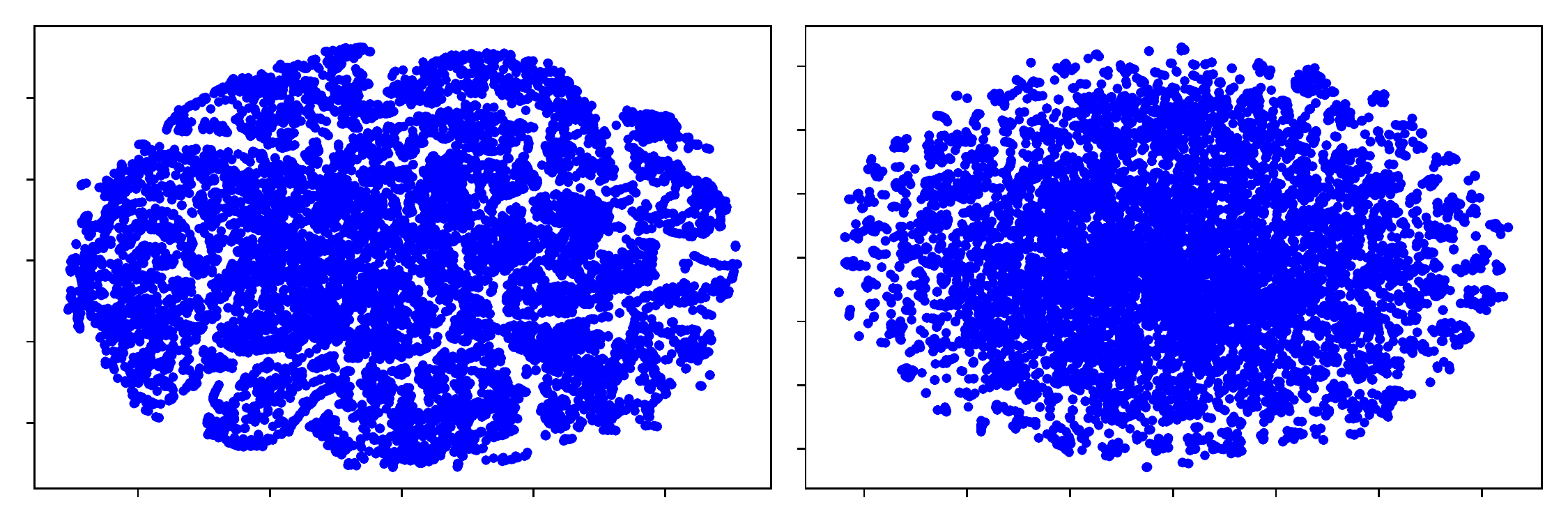}
    \caption{The t-SNE visualization of event representations from encoders without and with contrastive learning.}
    \label{fig:tsne}
\end{figure}

In Figure~\ref{fig:tsne}, we present 2-dimensional t-SNE visualizations of the representation space obtained from queries without and with contrastive learning. As demonstrated in the left part of the figure, without contrastive learning, the model encodes queries into a smaller space with more collapses. As a comparison, the addition of contrastive learning expands the embedding space with better alignment and uniformity.

\section{Event Filter Rules}\label{sec:event_fi}
In this stage, we introduce the filtering criteria for headlines in the coarse filter phase. Five types of headlines will be excluded, namely: non-event headlines, missing important components related to the event, containing multiple events, irregular syntax, and interrogative sentences. We provide examples of each of these five types in Figure~\ref{tab:filter_example}.

\end{CJK*}
\end{document}